\documentstyle[preprint,prb,aps,epsf,eqsecnum]{revtex}
\begin{document}
\draft
\tightenlines
\title{Magnetic Properties of the low dimensional spin system
(VO)$_2$P$_2$O$_{7}$: ESR and susceptibility }

\author{A. V. Prokofiev, F. B\"ullesfeld, W. Assmus, H. Schwenk, D. Wichert,
 U. L\"ow, B. L\"uthi}
\address
{Physikalisches Institut, Universit\"at Frankfurt,
 Robert Mayer Str. 2-4, D-60054 Frankfurt }

\maketitle

\begin{abstract}
\begin{center}
\parbox{14cm}{Experimental results on magnetic resonance (ESR)
and magnetic susceptibility are given for single crystalline
(VO)$_2$P$_2$O$_{7}$. The crystal growth procedure is briefly
discussed .The susceptibility is interpreted numerically
using a model with alternating spin chains. We determine $J$=51 K and $\delta$=0.2.
Furthermore we find a spin gap of $\approx6$meV from our ESR
measurements. Using elastic constants no indication of a phase
transition forcing the dimerization is seen below 300 K.  }
\end{center}
\end{abstract}

\newpage
\section{Introduction}
\label{I}

The interest in low dimensional spin systems is motivated by the occurrence
of many different ground states and of excitations with spin gaps. Prominent
examples are compounds with spin chains exhibiting Spin-Peierls transitions,
or dimerisation due to nearest and to
next nearest exchange interactions or Spin 1 systems.
In addition spin ladders show also unexpected results\cite{Dag}.
\\

The magnetism of the compound (VO)$_2$P$_2$O$_7$ (abbreviated VOPO)
experienced quite different interpretations. The first susceptibility
measurements on polycrystalline material  \cite{Joh}
were interpreted with an alternating chain model and a two leg
ladder model  \cite{BaRi}. Subsequent inelastic neutron scattering
data indicating a spin gap of about  50K were used as indication
of a ladder model  \cite{Ecc}. Recently neutron scattering
experiments on an arrangement of many (200) tiny single crystals
(dimensions $< 1$mm$^3$) gave evidence of strong dispersion in
the b-direction of this orthorhombic compound
(see Fig. 1) thereby eliminating the two leg ladder model
and installing again the alternating chain model\cite{Gar}.
Our previous susceptibility and magnetic resonance work
on polycrystalline VOPO was interpreted with the two leg
ladder model \cite{Sch,Sch1}.
\\

Recently we succeeded in growing large single
crystals in excess of 1mm$^3$. We present in this
paper a brief description of the crystal growth
procedure and we present magnetic resonance (ESR) and
susceptibility data for these single crystals as a function
of temperature and magnetic field. We give an interpretation
of our results using the alternating chain model \cite{Gar}.

\section{Experiment}
\label{II}

The growth of single crystals of VOPO is complicated by two
features of this compound. The first is the strong
sensitivity of the oxidation state of
Vanadium (and of the stability of VOPO) to the oxygen content
in the growth atmosphere. The second is the tendency of
VOPO-melt to a glass formation during cooling due to the high
viscosity. Therefore the growth has to be carried out
in an atmosphere with strictly controlled oxygen content
and with a very low growth rate.

VOPO powder for the growth was prepared by thermal decomposition
of the precursor (VO)HPO$\cdot5H_2$O in argon flow at
$700^ \circ C$. The precursor
was synthesized according to Centi et al.\cite{Cen} .
Single crystals were
grown by pulling with the velocity of 2mm/day from the melt
with simultaneous cooling of the melt with the
rate 4-8 K  per day. Single crystals with sizes
up to $10\times3\times3 $ mm$^3$ have been grown in a week.
The details of the growth technique will
be published elsewhere \cite{Pro}.

Growth atmosphere with various oxygen contents were used.
The latter varied in the limits 0.2-0.6 vol$\%$ for different runs.
This concentration regions provide the stability
of the phase (VO)$_2$P$_2$O$_{7+x}$, which has a homogeneity
range in oxygen content \cite{Lop}. This, however,
requires the determination of the oxygen content
(or the Vanadium oxidation state) in the growth crystals.
Thermogravimetric analysis based on the reactions of either
oxidation of (VO)$_2$P$_2$O$_{7+x}$ up to VPO$_5$ or mild reduction
in vacuum to stoichiometric composition
(VO)$_2$P$_2$O$_7$ showed that the Vanadium oxidation
state in these conditions varied in the limits $4.0-4.4$.
\\

We measured the susceptibility either with a vibrating sample
magnetometer or as magnetization in a constant field of $1-2T$.
Both methods gave the same results.
An absolute calibration was performed with a
Faraday balance. For the
magnetic resonances we used various diodes in the frequency
range $130-300$ GHz in combination with frequency multipliers
\cite{Exp}.

\section{Results and Discussion}
\label{III}
\subsection{Resonance}
\label{IIIA}

In Fig. 2 we first present magnetic resonance results in the
frequency region 134 - 288 GHz. In the three crystallographic
directions a,b,c we find Zeeman split lines with g-factors
$g_a = 1.937$, $g_b = g_c = 1.984$. Comparing these g-factors
with previous results on polycrystalline specimens \cite{Sch}
($g^\parallel = 1.94$, $g^ \perp = 1.98$) we note the
good agreement of the latter
values with the present ones.
This gives a clear indication of the successful
averaging procedure to obtain the lineshape
in the polycrystalline material.
In addition we see some small resonances in the single crystal
similar to the resonance form seen in Ref.\onlinecite{Sch}.
Details will be discussed elsewhere.
In the polycrystalline sample
the total absorption of the ESR line could be fitted to a singlet
triplet energy gap of 150 K (see Ref.\onlinecite{Sch1}). We have assumed 
that this
temperature behaviour could arise from transitions at $k$=0. In the
polycrystalline material an energy gap of this size was observed in
inelastic neutron scattering\cite{Ecc}.
\\

For the single crystal resonances of Fig. 2 arise from excited triplet excitations too,
as shown in Fig. 3. Here the total absorption strength for a
resonance of 134 GHz is shown as a function of temperature.
The full line indicates the calculated intensity  of  an excitation
within the triplet with a fitted singlet-triplet gap of 1374 GHz = 67K.
This gives a satisfactory description of the absorption.
This energy gap fits to the upper mode of the measured
excitation spectra \cite{Gar} . The small dispersion in the c-direction
gives an enhanced density of states for such a resonance.
The dotted curve gives an analogous fit for an excitation
of 849GHz = 36K corresponding to the lower observed branch
of Ref.\onlinecite{Gar}. This curve does not fit the experimental results
at all. An alternating chain model, discussed in section \ref{IIIC}, with $J=51$K and $\delta$=0.2
gives a gap of 58K ( see Ref. \onlinecite{Uhr}) close 
to the observed one of 67K.
More detailed information of the excitation spectra
are needed to reach a final conclusion for the observed
triplet excitation.

\subsection{Susceptibility}
\label{IIIB}

In Fig. 4a we present the susceptibility.
The curves along the different
axis are slightly different. They all have a maximum at
$74 \pm 2K$
close to the broad maximum  observed for the polycrystalline
specimens \cite{Joh,Sch}.

The remarkable feature of the single crystal result is however the
almost total absence of a low temperature tail compared to the
previous polycrystal work (compare e.g. the inset of Fig.1
in Ref. \onlinecite{Sch}
with our
results in Fig. 4a). In the polycrystalline material we
could fit the defect part of the susceptibility to a $T^{-0.54}$ law.
In the single crystal measurements it is not possible to fit a
law to the defect part of the susceptibility because the temperature
region is too small.
This defect part is due to chain ends or
paramagnetic $V^{4+}$ ions which are isolated by nonmagnetic
$V^{5+}$ ions.
In the inset of Fig. 4a we show 2T and 10T ($\chi=M/B$) susceptibility data.
For the high field data the low temperature tail is completely suppressed as expected.
\\

The difference in the temperature dependence of the susceptibility for
the different crystallographic directions is mainly due to the
different g-factors. The g-factor for the $<100>$ direction is
smaller than the g-factors for the other directions as discussed
above. This is similar to the findings of
Ref. \onlinecite{Sch} and to the ESR measurements as pointed out above.
The absolute value of the maximum of the susceptibility
is about $2* 10^{-3}$ emu/molV. These values are slightly higher than
the maximum value of $\chi(T)$ in
Ref. \onlinecite{Joh}  because in our single crystal
 samples the valence state of the V-ions is closer to 4+. There
are more magnetic ions that contribute to the susceptibility of the chains.
\\

To fit a Curie law to the susceptibility data one had to
measure to higher temperatures than 230K. The Curie-Weiss law
for the $<010>$ direction gives an approximate Curie constant
of C = .374 emu/molV and an upper limit of a Curie Weiss
temperature of $-70K$. C gives an effective  Spin per ion of $S = 1/2$
if we take the g-factor of $g = 1.984$ from above.
So the susceptibility gives also a valence of about 4.0
for the V-ions similar to the oxidation reactions.

\subsection{Discussion of the susceptibility}
\label{IIIC}

We now analyse the susceptibility measurement quantitatively
in the framework of the spin 1/2 Heisenberg model
with alternating couplings given by the Hamiltonian

\begin{equation}
H = 2 J \sum_{i = 1}^{N} \left\{(1+\delta  (-1)^i) \vec S_{i} \vec S_{i + 1}
 \right\} .
\end{equation}

We determined $J$
by comparing the experimentally found temperature
$T_{max}= (74 \pm 2)K $
where $\chi$ has its maximum
with the theoretical result for $T_{max}/J$
from exact diagonalization. For a successful application of this
approach see Ref. \onlinecite{Fab}.
Subsequently we used $J(\delta)$ to calculate
the maximum $\chi_{max}=\chi(T_{max})$ of the susceptibility in emu/molV
as a function of $\delta$. The results are shown in Fig. 5a and 5b
respectively.
From the experimental value $\chi_{max}=2.07*10^{-3}$ emu/molV
with $g=2$ and with $T_{max}$=74K we find $\delta=0.2  $ and $J=51K  $.
The calculated susceptibility for these parameters 
together with
the experimental result
in the $<010>$ direction are shown in Fig. 4b. Note that no defect part is
subtracted as mentioned in section \ref{IIIB}. This explains the less
good agreement for $T < T_{max}$.
\\

The maxima used in Fig. 5 are obtained
by exact diagonalization
of the complete Hamiltonian for up to 16 spins and
$\delta = 0.0 ... 1.0$.
Finite size effects do not play a role in the
temperature range of $T_{max}$. This can be
seen from Fig 5a and b where the results
for chains with N=14 and 16 spins fall together.
\\

We point out that the influence of an additional frustrating interaction
$ 2 J \sum_{i = 1}^{N} \alpha S_{i} S_{i + 2} $ shifts
the maximum of the susceptibility to smaller temperatures $T/J$
and consequently for the same experimental $T_{max}$
we obtain larger values for the coupling J.
The absolute value of $\chi$
is lowered both by the frustration,
and by the scaling factor $1/J$ (see Ref. \onlinecite{Low}).
\\

In Fig. 6 we show the elastic constant $c_{11}$ as a function
of temperature. The absolute value of the sound velocity is 5097 m/s at 4.2K.
No indication of a phase transition is observed for
$T\leq300$K. A discussion of the different elastic
constants will be given later.
\\

In conclusion we have performed susceptibility and ESR measurements
on single crystalline samples. Based on a model of a dimerized chain we
calculated J=51K and $\delta$=0.2. Our ESR-measurements show a singlet-triplet
gap of 67 K in good agreement to the upper gap as measured in inelastic neutron
scattering. The elastic constant $c_{11}$ shows that the dimerization is 
not due to a phase transition. 
Further work has to investigate the details of the ESR-Signal 
and to
estimate the influence of possible interchain exchange interactions
(as seen in Ref.\onlinecite{Gar}).

\vskip 3cm
Acknowledgement\\

This research was supported in part by SFB 252. We thank
G. Bouzerar, K. Fabricius, A. P. Kampf and G.S. Uhrig for enlightening discussions.

\newpage

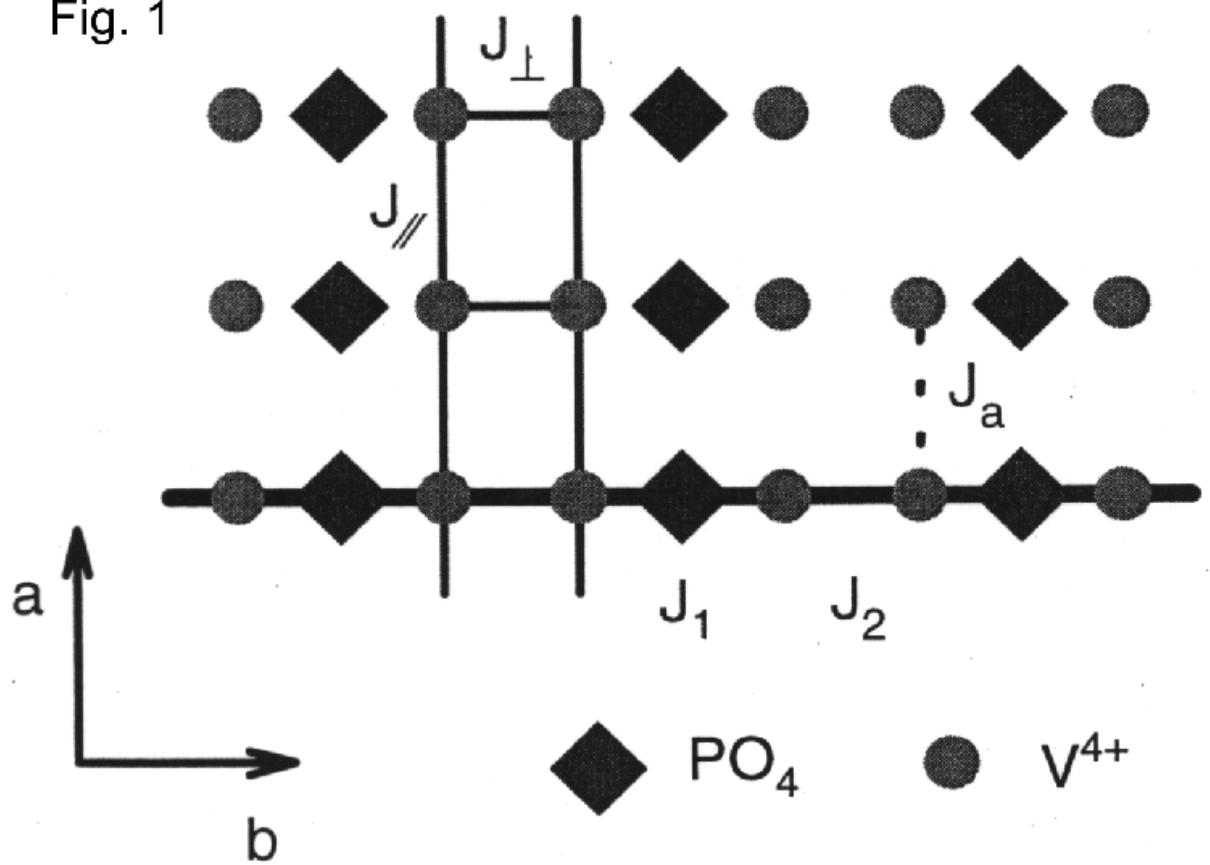
\noindent{Fig. 1:\\
Schematic description of the structure and magnetic
interactions in VOPO, adapted from Ref. \onlinecite{Gar}\\
\\
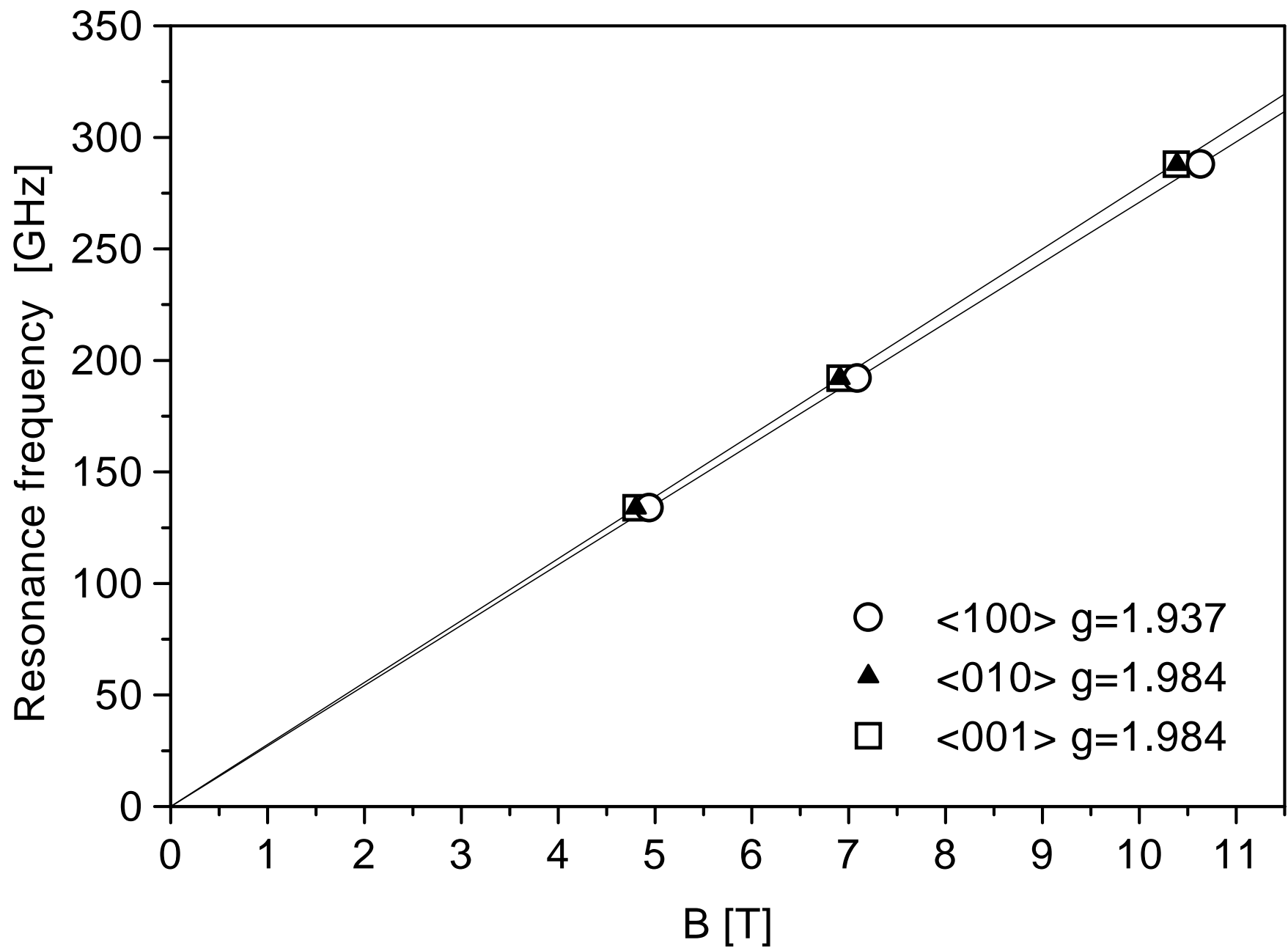
Fig. 2:\\
Resonance frequencies versus magnetic field for
the different crystallographic directions.\\
\\
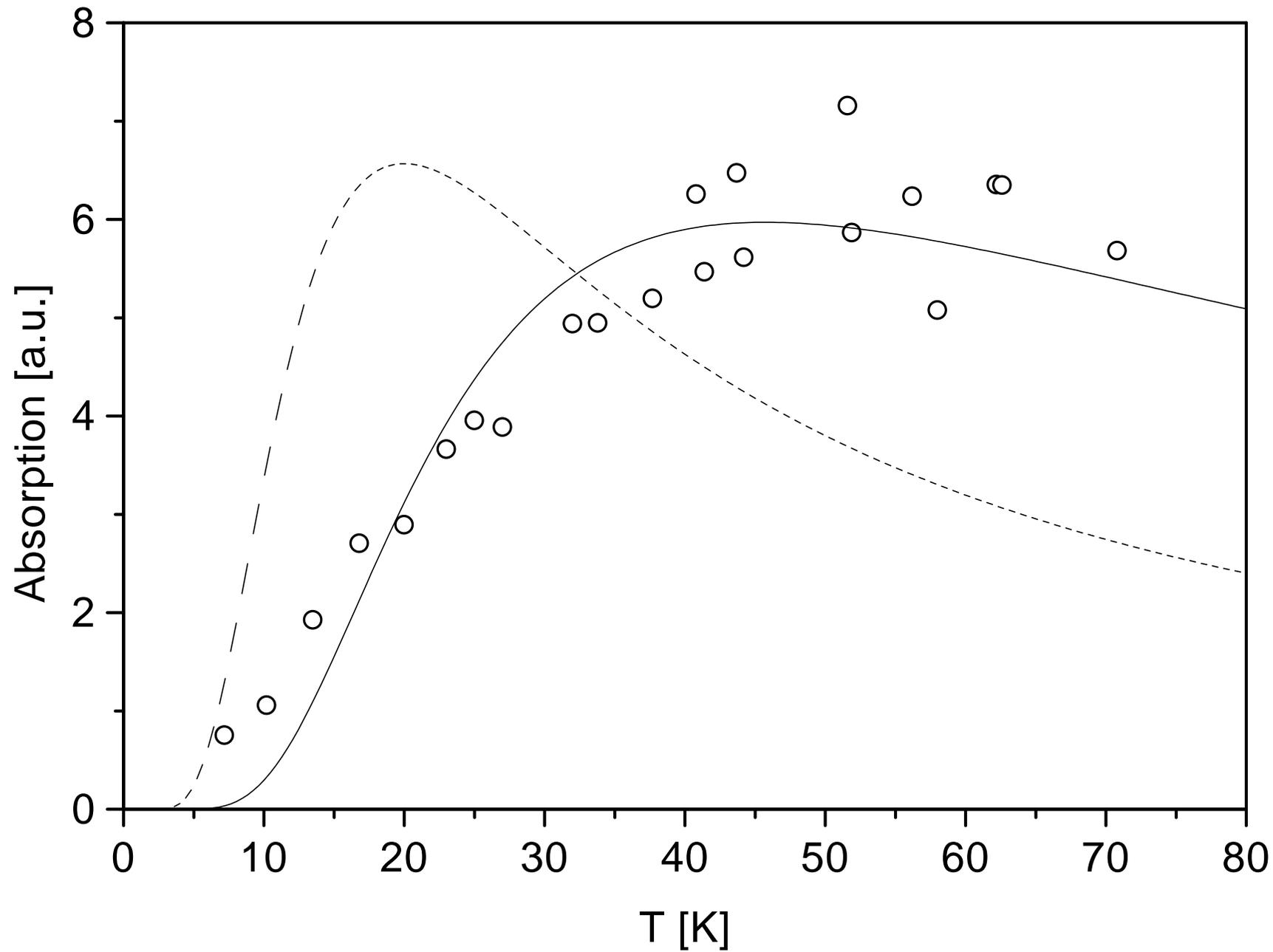
Fig. 3:\\
Absorption strength of the 134 GHz line as a function
of temperature. The full and dotted lines are fits for
a resonance within the triplet state as explained in the text.\\
\\
Fig. 4a:\\
Temperature dependence of magnetic susceptibility in
VOPO for the different crystal directions. In the inset we give
the low temperature susceptibility ($\chi$=$M/B$) for 2T and 10T.\\
\\
Fig. 4b:\\
Temperature dependence of magnetic susceptibility in
VOPO for the $<010>$ direction (open circles) and calculated
susceptibility with $J$=51K and $\delta$=0.2 (full line).\\
\\
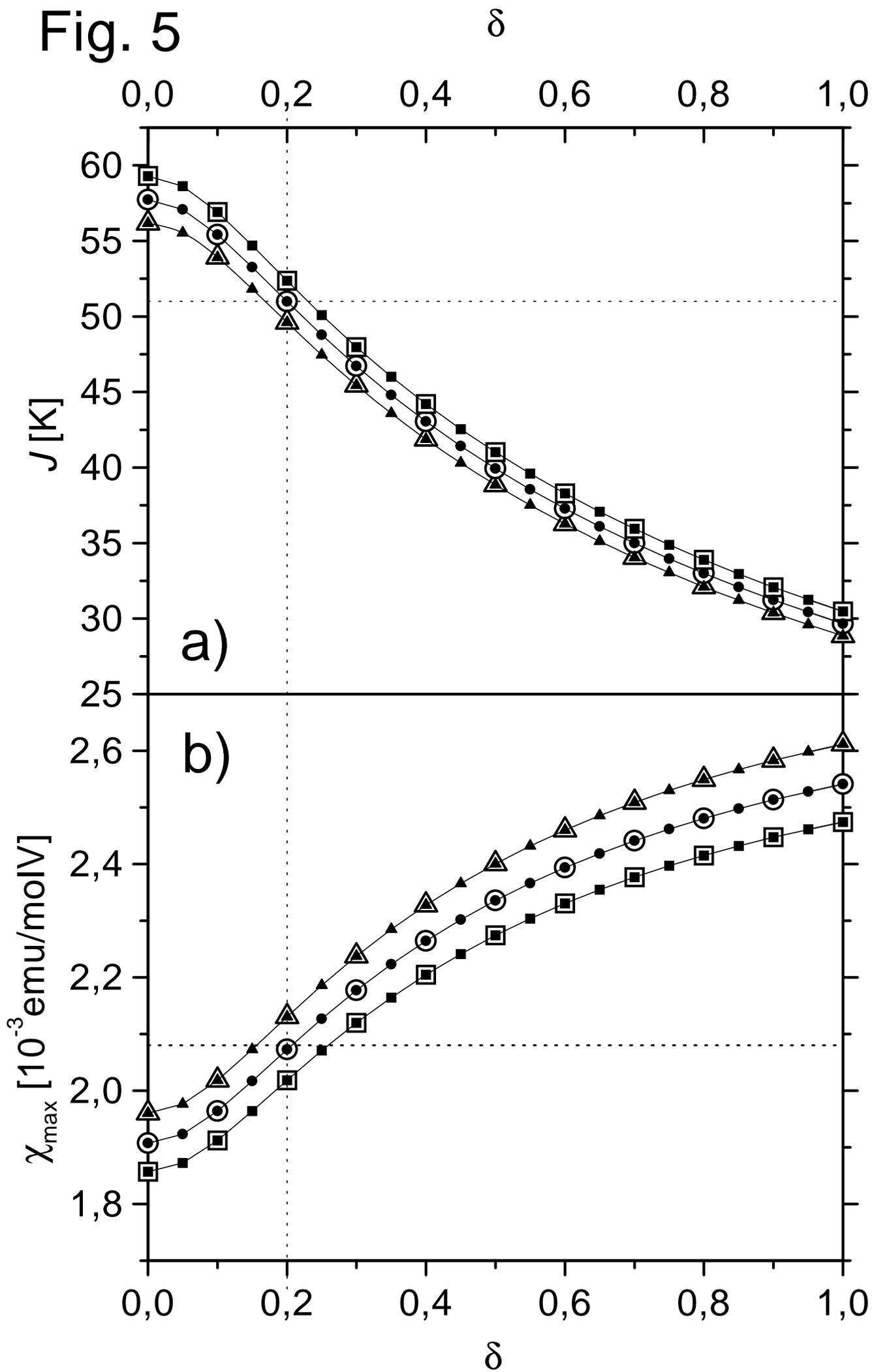
Fig. 5:\\
a) Antiferromagnetic exchange coupling $J(\delta)$.\\
b) Maximum susceptibility $\chi_{max}(\delta)$ normalized to $g=2$.\\
$T_{max}$=72K (squares),74K(circles),76K(triangles).
Small filled symbols for $N=$14, large open symbols for $N=$16.\\
\\
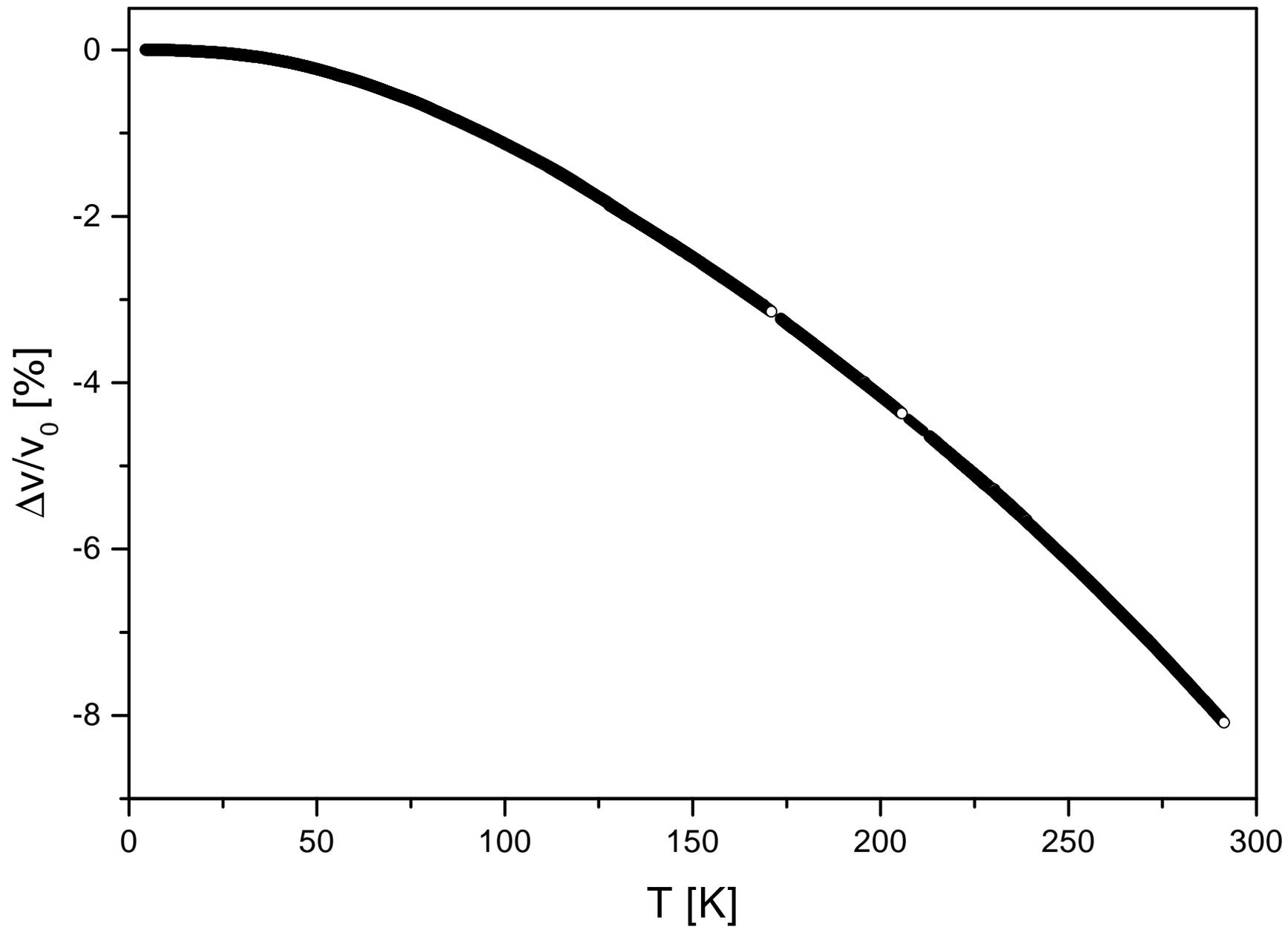
Fig. 6:\\
Relative sound velocity for the $c_{11}$-mode.\\
Absolute sound velocity for the $c_{11}$-mode: $v$=5097m/s at $T=$4.2K.}


\begin{references}

\bibitem{Dag} E.Dagotto, T.M.Rice, Science 271, 618 (1996)
\bibitem{Joh} D.C.Johnston, J.W.Johnson, D.P.Gashorn, A.J.Jacobson, Phys.Rev.B35, 219(1987)
\bibitem{BaRi} T.Barnes and J.Riera, Phys.Rev.B50, 6817(1994)
\bibitem{Ecc} R.S.Eccleston, T.Barnes, J.Brody, J.W.Johnson, Phys.Rev.Lett. 73, 2626(1994).
\bibitem{Gar} A.W.Garrett et al., Phys.Rev.Lett. 79,745(1997)
\bibitem{Sch} H.Schwenk, M.Sieling, D.K\"onig, W.Palme, S.A.Zvyagin, B.L\"uthi, R.S.Eccleston Solid State Comm. 100, 381(1996)
\bibitem{Sch1} H.Schwenk, D.K\"onig, M.Sieling, S.Schmidt, W.Palme, B.L\"uthi, S.A.Zvyagin, R.S.Eccleston, M.Azuma, M.Takano Physica B 237-238, 115(1997)
\bibitem{Cen} G.Centi, F.Trifiro, G.Poli, Appl.Catal. 19,225(1985).
\bibitem{Pro} A.V.Prokofiev, F.B\"ullesfeld, W.Assmus, Crystal Research and Tech-.(to be publ.).
\bibitem{Lop} M.Lopez, Granados, J.C.Conesa, M.Fernandes-Garcia, J.Catal. 141,671(1993)
\bibitem{Exp} W.Palme et al., Z.Phys.B92, 1(1993)
\bibitem{Low} U.L\"ow to be published
\bibitem{Fab} K.Fabricius, A.Kl\"umper, U.L\"ow, B.B\"uchner, T.Lorenz,
              G.Dhalenne, A.Revcolevschi, to appear in
              Phys.Rev.B57,(1998)
\bibitem{Uhr} G.S.Uhrig, H.J.Schulz, Phys.Rev.B54, 54(1996)

\end{references}
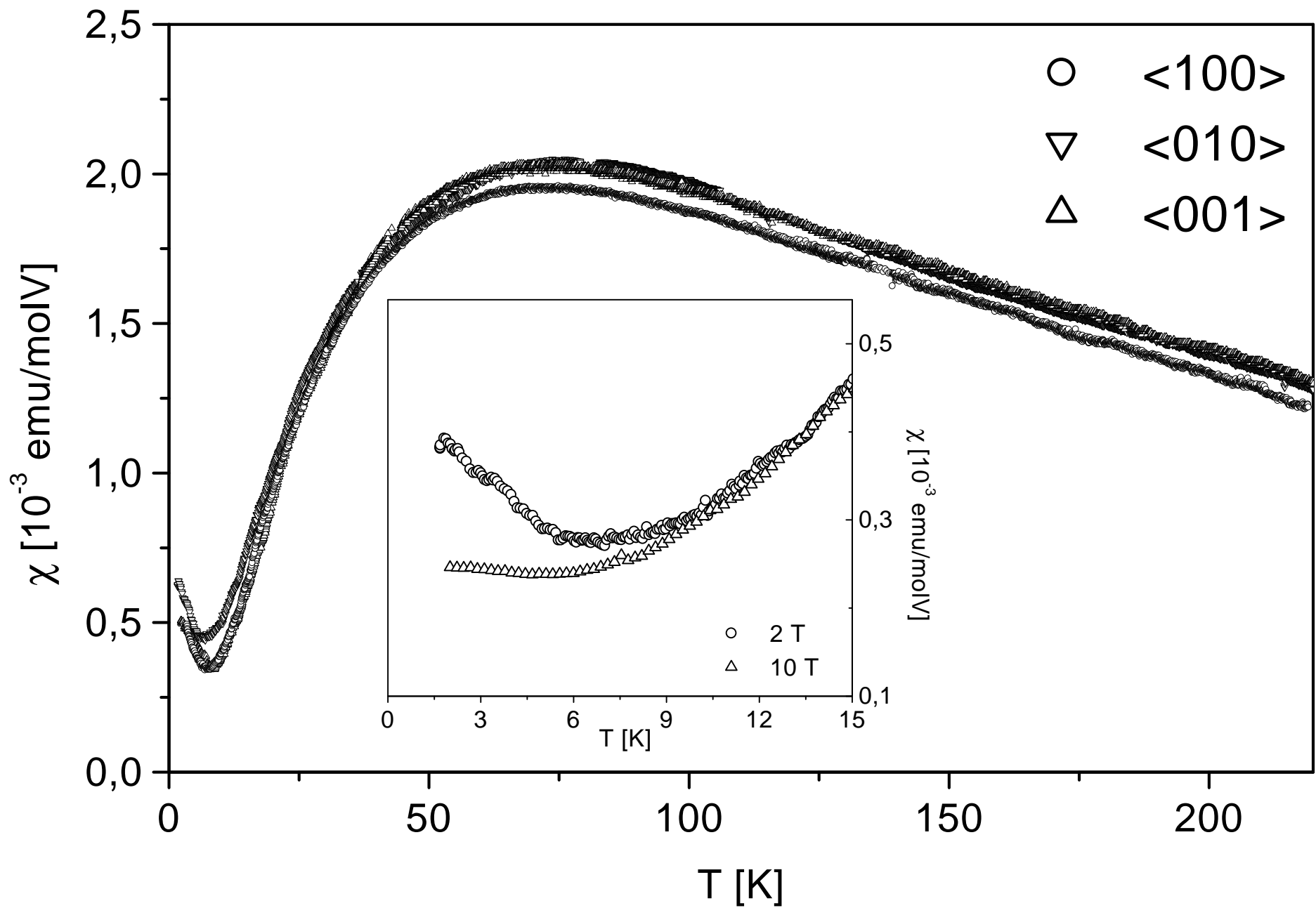
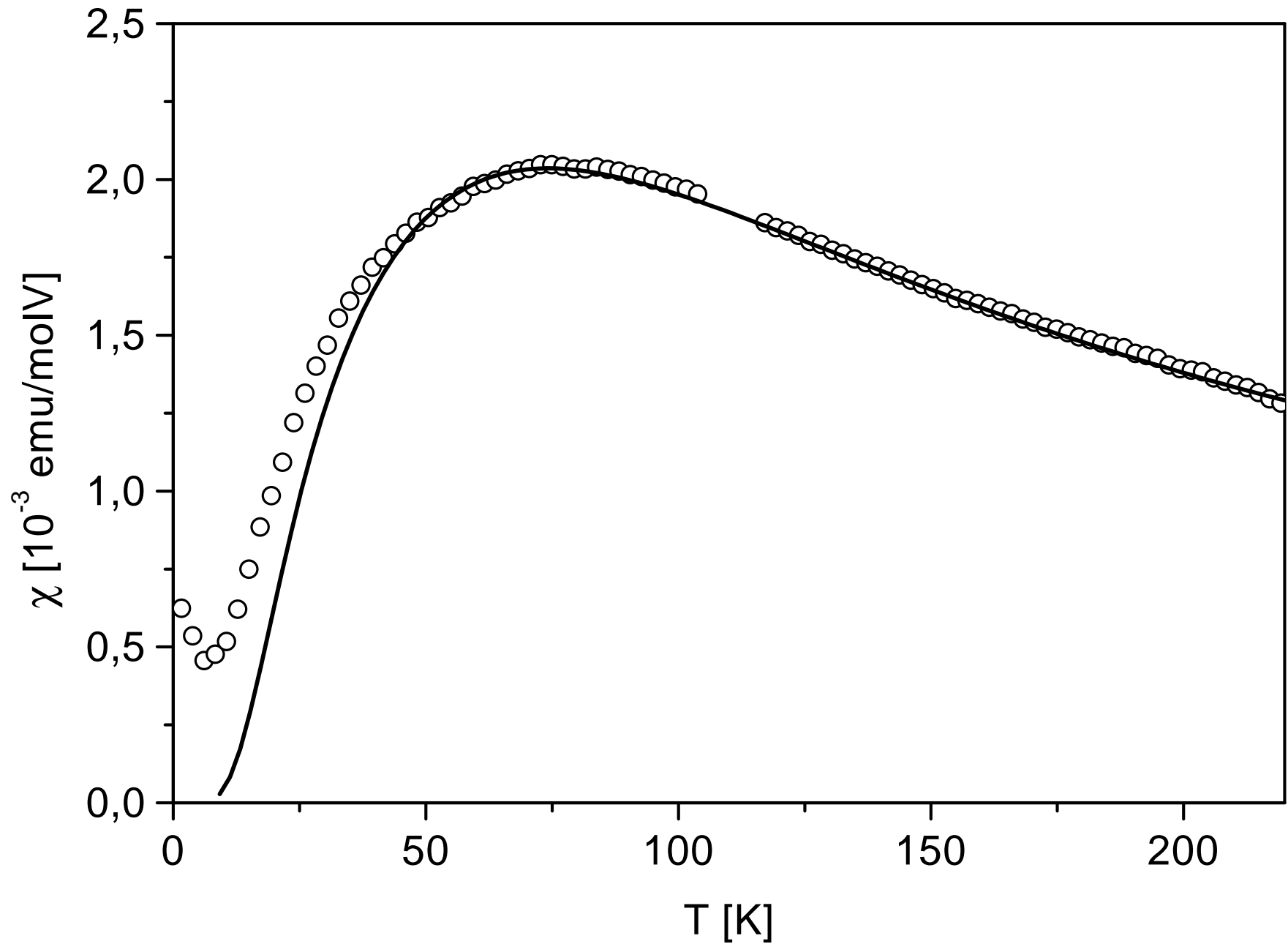
\end{document}